\documentclass[12pt]{article}

\catcode`@=12
\begin{document}
\thispagestyle{empty}
\begin{flushright} 
UCRHEP-T342\\ 
August 2002\
\end{flushright}
\vspace{0.5in}
\begin{center}
{\LARGE	\bf The All-Purpose Neutrino Mass Matrix\\}
\vspace{1.5in}
{\bf Ernest Ma\\}
\vspace{0.2in}
{\sl Physics Department, University of California, Riverside, 
California 92521\\}
\vspace{1.5in}
\end{center}
\begin{abstract}\
A four-parameter Majorana neutrino mass matrix is proposed, which is 
exactly diagonalizable with two mixing angles, one of which is $\pi/4$ 
and the other large but less than $\pi/4$.  The form of this mass matrix 
allows seven realizations of the present data on atmospheric and solar 
neutrino oscillations: three have the normal hierarchy, two have the 
inverse hierarchy, and two have three nearly degenerate masses.  The 
possible inclusion of CP violation is also discussed.
\end{abstract}
\newpage
\baselineskip 24pt

With the recent addition of the SNO (Sudbury Neutrino Observatory) 
neutral-current data \cite{sno}, the overall picture of solar neutrino 
oscillations \cite{sol} is becoming quite clear.  Together with the 
well-established atmospheric neutrino data \cite{atm}, the neutrino 
mixing matrix is now determined to a very good first approximation by
\begin{equation}
\pmatrix {\nu_1 \cr \nu_2 \cr \nu_3} = \pmatrix {\cos \theta & \sin \theta/
\sqrt 2 & \sin \theta/\sqrt 2 \cr -\sin \theta & \cos \theta/\sqrt 2 & 
\cos \theta/\sqrt 2 \cr 0 & -1/\sqrt 2 & 1/\sqrt 2} \pmatrix {\nu_e \cr 
\nu_\mu \cr \nu_\tau},
\end{equation}
where $\nu_{1,2,3}$ are Majorana neutrino mass eigenstates.  In the above, 
$\sin^2 2 \theta_{atm} = 1$ is already assumed and $\theta$ is the solar 
mixing angle which is now known to be large but not maximal \cite{many}, i.e. 
$\tan^2 \theta \simeq 0.4$.  The $U_{e3}$ entry has been assumed zero but 
it is only required to be small \cite{react}, i.e. $|U_{e3}| < 0.16$. 
A nonzero $U_{e3}$ with a possible complex phase for CP violation will be 
discussed later.

Denoting the masses of $\nu_{1,2,3}$ as $m_{1,2,3}$, the solar neutrino data 
\cite{sno,sol} require that $m_2^2 > m_1^2$ with $\theta < \pi/4$, and in 
the case of the favored large-mixing-angle solution \cite{many},
\begin{equation}
\Delta m^2_{sol} = m_2^2 - m_1^2 \simeq 5 \times 10^{-5}~{\rm eV}^2.
\end{equation}
The atmospheric neutrino data \cite{atm} require
\begin{equation}
|m_3^2 - m_{1,2}^2| \simeq 2.5 \times 10^{-3}~{\rm eV}^2,
\end{equation}
without deciding whether $m_3^2 > m_{1,2}^2$ or $m_3^2 < m_{1,2}^2$.

The big question now is what the neutrino mass matrix itself should look like. 
It is proposed here that it should be of the form
\begin{equation}
{\cal M}_\nu = \pmatrix {a+2b+2c & d & d \cr d & b & a+b \cr d & a+b & b}.
\end{equation}
This matrix has 4 parameters which are assumed first to be all real.  In that 
case, it is $exactly$ diagonalized by Eq.~(1) with its eigenvalues $exactly$ 
given by (I):
\begin{eqnarray}
m_1 &=& a+2b+c - \sqrt {c^2 + 2d^2}, \\ 
m_2 &=& a+2b+c + \sqrt {c^2 + 2d^2}, \\ 
m_3 &=& -a,
\end{eqnarray}
with
\begin{equation}
\tan \theta = {\sqrt {2} d \over \sqrt {c^2 + 2d^2} - c},
\end{equation}
for $a+2b+c > 0$, $c < 0$, and (II):
\begin{eqnarray}
m_1 &=& a+2b+c + \sqrt {c^2 + 2d^2}, \\ 
m_2 &=& a+2b+c - \sqrt {c^2 + 2d^2}, \\ 
m_3 &=& -a,
\end{eqnarray}
with
\begin{equation}
\tan \theta = {\sqrt {c^2 + 2d^2} - c \over \sqrt {2} d},
\end{equation}
for $a+2b+c < 0$, $c > 0$.

Note that $\theta$ depends only on the ratio $d/c$, which must be of order 
unity.  This shows the advantage for adopting the parametrization of 
Eq.~(4).  The constraints of Eqs.~(2) and (3) are then realized by the 
following 7 different conditions on $a$, $b$, and $c$.

(1) $||a+2b+c|-\sqrt{c^2+2d^2}| << |a+2b+c| << |a|$, ~~i.e. $|m_1| << |m_2| << 
|m_3|$.

(2) $\sqrt{c^2+2d^2} << |a+2b+c| << |a|$, ~~i.e. $|m_1| \simeq |m_2| << |m_3|$.

(3) $|a+2b+c| << \sqrt{c^2+2d^2} << |a|$, ~~i.e. $|m_1| \simeq |m_2| << |m_3|$.

(4) $|a|,~\sqrt{c^2+2d^2} << |a+2b+c|$, ~~i.e. $|m_3| << |m_1| \simeq |m_2|$.

(5) $|a|,~|a+2b+c| << \sqrt{c^2+2d^2}$, ~~i.e. $|m_3| << |m_1| \simeq |m_2|$.

(6) $\sqrt{c^2+2d^2} << ||a+2b+c|-|a|| << |a|$, ~~i.e. $|m_1| \simeq |m_2| 
\simeq |m_3|$.

(7) $|a+2b+c| << \sqrt{c^2+2d^2} \simeq |a|$, ~~i.e. $|m_1| \simeq |m_2| 
\simeq |m_3|$.

\noindent Cases (1) to (3) have the normal hierarchy. Cases (4) and (5) have 
the inverse hierarchy. Cases (6) and (7) have 3 nearly degenerate masses. 
The versatility of Eq.~(4) has clearly been demonstrated.

The above 7 cases encompass all models of the neutrino mass matrix that 
have ever been proposed which also satisfy Eq.~(1).  They are also very 
useful for discussing the possibility of neutrinoless double beta ($\beta 
\beta_{0\nu}$) decay in the context of neutrino oscillations \cite{bbmass}.  
The effective mass $m_0$ measured in $\beta \beta_{0\nu}$ decay is 
$|a+2b+2c|$.  However, neutrino oscillations constrain $|a+2b+c|$ and 
$\sqrt{c^2+2d^2}$, as well as $|d/c|$.  Using
\begin{equation}
|a+2b+2c| = ||a+2b+c| \pm |c|| = ||a+2b+c| \pm \cos 2 \theta \sqrt{c^2+2d^2}|,
\end{equation}
the following conditions on $m_0$ are easily obtained:
\begin{eqnarray}
(1) && m_0 \simeq \sin^2 \theta |m_2| \simeq \sin^2 \theta \sqrt 
{\Delta m^2_{sol}}, \\ 
(2) && m_0 \simeq |m_{1,2}| << \sqrt {\Delta m^2_{atm}}, \\ 
(3) && m_0 \simeq  \cos 2 \theta |m_{1,2}| << \cos 2 \theta \sqrt 
{\Delta m^2_{atm}}, \\ 
(4) && m_0 \simeq \sqrt {\Delta m^2_{atm}}, \\ 
(5) && m_0 \simeq \cos 2 \theta \sqrt {\Delta m^2_{atm}}, \\ 
(6) && m_0 \simeq |m_{1,2,3}|, \\ 
(7) && m_0 \simeq \cos 2 \theta |m_{1,2,3}|.
\end{eqnarray}
If $m_0$ is measured \cite{klapdor} to be significantly larger than 0.05 eV, 
then only Cases (6) and (7) are allowed.  However, as Eqs.~(19) and (20) 
show, the true mass of the three neutrinos is still subject to a two-fold 
ambiguity, which is a well-known result.

The underlying symmetry of Eq.~(4) which results in $U_{e3} = 0$ is its 
invariance under the interchange of $\nu_\mu$ and $\nu_\tau$.  Its mass 
eigenstates are then separated according to whether they are even 
$(\nu_{1,2})$ or odd $(\nu_3)$ under this interchange, as shown by Eq.~(1). 
To obtain $U_{e3} \neq 0$, this symmetry has to be broken.  One interesting 
possibility is to rewrite Eq.~(4) as
\begin{equation}
{\cal M}_\nu = \pmatrix {a+2b+2c & d & d^* \cr d & b & a+b \cr d^* & a+b & b},
\end{equation}
where $a,b,c$ are real but $d$ is complex.  This reduces to Eq.~(4) if 
$Im d = 0$, but if $Im d \neq 0$, then $U_{e3} \neq 0$.

To obtain $U_{e3}$ in a general way, first rotate to the basis spanned by 
$\nu_e, (\nu_\mu+\nu_\tau)/\sqrt 2$, and $(\nu_\tau-\nu_\mu)/\sqrt 2$, i.e.
\begin{equation}
{\cal M}_\nu = \pmatrix {a+2b+2c & \sqrt2 Red & -\sqrt2 i Imd \cr \sqrt2 Red 
& a+2b & 0 \cr -\sqrt 2 i Imd & 0 & -a}
\end{equation} 
Whereas ${\cal M}_\nu$ is diagonalized by
\begin{equation}
U {\cal M}_\nu U^T = \pmatrix {m_1 & 0 & 0 \cr 0 & m_2 & 0 \cr 0 & 0 & m_3},
\end{equation}
${\cal M}_\nu {\cal M}_\nu^\dagger$ is diagonalized by
\begin{equation}
U ({\cal M}_\nu {\cal M}_\nu^\dagger) U^\dagger = \pmatrix {|m_1|^2 & 0 & 0 
\cr 0 & |m_2|^2 & 0 \cr 0 & 0 & |m_3|^2}.
\end{equation}
Here
\begin{equation}
{\cal M}_\nu {\cal M}_\nu^\dagger = \pmatrix {(a+2b+2c)^2 + 2|d|^2 & 
2 \sqrt2 (a+2b+c) Red & 2 \sqrt2 i (a+b+c) Imd \cr 2 \sqrt2 (a+2b+c) Red & 
(a+2b)^2 + 2(Red)^2 & 2 i Red Imd \cr -2 \sqrt2 i (a+b+c) Imd & -2 i Red Imd 
& a^2 + 2(Imd)^2}.
\end{equation}
To obtain $U_{e3}$ for small $Imd$, consider the matrix
\begin{equation}
A = {\cal M}_\nu {\cal M}_\nu^\dagger - [a^2 + 2(Imd)^2] I,
\end{equation}
where $I$ is the identity matrix. Now $A$ is diagonalized by $U$ as well and 
$U_{e3}$ is simply given by
\begin{equation}
U_{e3} \simeq {A_{e3} \over A_{ee}} = {2 \sqrt 2 i (a+b+c) Imd \over 
(a+2b+2c)^2 - a^2 + 2 (Red)^2}
\end{equation}
to a very good approximation and leads to
\begin{eqnarray}
(1), (2), (3) && U_{e3} \simeq {-\sqrt 2 i Imd \over a}, \\ 
(4), (6) && U_{e3} \simeq {i Imd \over \sqrt 2 b}, \\ 
(5) && U_{e3} \simeq {\sqrt2 i Imd \over c}, \\ 
(7) && U_{e3} \simeq {\sqrt2 i (a+c) Imd \over c^2 - a^2 + 2(Red)^2}.
\end{eqnarray}
In all cases, the magnitude of $U_{e3}$ can be as large as the present 
experimental limit \cite{react} of 0.16 and its phase is $\pm \pi/2$. 
Thus the CP violating effect in neutrino oscillations is predicted 
to be maximal by Eq.~(21), which is a very desirable scenario for future 
long-baseline neutrino experiments.

The above analysis shows that for $U_{e3} = 0$ and $\sin^2 2 \theta_{atm} = 1$,
the seven cases considered cover all possible patterns of the $3 \times 3$ 
Majorana neutrino mass matrix, as indicated by present atmospheric and solar 
neutrino data.  Any successful model should predict Eq.~(4) at least as a 
first approximation.  One such example already exists \cite{a4}, where 
$b=c=d=0$ corresponds to the non-Abelian discrete symmetry $A_4$, i.e. the 
finite group of the rotations of a regular tetrahedron.  This leads to 
Case (6), i.e. three nearly degenerate masses, with the common mass equal 
to that measured in $\beta \beta_{0\nu}$ decay.  It has also been shown 
recently \cite{bmv} that starting with this pattern, the correct mass 
matrix, i.e. Eq.~(21) with the complex phase in the right place, is 
automatically obtained with the most general application of radiative 
corrections.  In particular, if soft supersymmetry breaking is assumed to 
be the origin of these radiative corrections, then the neutrino mass matrix 
is correlated with flavor violation in the slepton sector, and may be tested 
in future collider experiments.

If $m_0$ turns out to be of order $\sqrt {\Delta m^2_{atm}}$ or less, then 
Cases (6) and (7) would be ruled out, and further progress in understanding 
the pattern of neutrino oscillations will depend on having a rationale for 
choosing one of the remaining 5 cases.  Consider for example Case (1) with 
$a+2b+c = \sqrt{c^2+2d^2}$ and $c=-d/2<0$, then Eq.~(4) becomes
\begin{equation}
{\cal M}_\nu = \pmatrix {d & d & d \cr d & b & -b+2d \cr d & -b+2d & b},
\end{equation}
which results in $m_1 = 0$, $m_2 = 3d$, $m_3 = 2b-2d$, and $\tan^2 \theta = 
0.5$.

As another example, consider Case (4) with $a=0$, $c=-d/2 < 0$, then Eq.~(4) 
becomes
\begin{equation}
{\cal M}_\nu = \pmatrix {2b-d & d & d \cr d & b & b \cr d & b & b},
\end{equation}
which results in $m_1 = 2b-2d$, $m_2=2b+d$, $m_3=0$, and $\tan^2 \theta = 
0.5$.  Thus $m_2^2-m_1^2 \simeq 12db$ and $m_{1,2}^2 - m_3^2 \simeq 4b^2$, 
where $d << b$ has been used.

Both examples are simple acceptable two-parameter solutions of the present 
atmospheric and solar neutrino data.  This shows the utility of Eq.~(4) as 
a way to find possible patterns which may even turn out to be supported by 
some symmetry.  As further examples, the case $c=0$ was discussed in Ref.~[10] 
and the case $a+2b+c=0$ was discussed in Ref.~[11].

In conclusion, recent experimental progress on neutrino oscillations points 
to a neutrino mixing matrix which can be understood in a systematic way 
in terms of an all-purpose neutrino mass matrix, i.e. Eq.~(4), and its 
simple extension, i.e. Eq.~(21), to allow for a nonzero and $imaginary$ 
$U_{e3}$, i.e. Eq.~(27).  Seven possible cases have been identified, each 
with a different prediction for $\beta \beta_{0\nu}$ decay, i.e. Eqs.~(14) 
to (20).  The present analysis should be useful as both a checklist and a 
roadmap for the further theoretical understanding of the origin of neutrino 
masses and mixing.\\[5pt]

This work was supported in part by the U.~S.~Department of Energy
under Grant No.~DE-FG03-94ER40837.

\newpage
\bibliographystyle{unsrt}

\end{document}